 \documentclass{article}
 \usepackage{graphicx}
 \usepackage{amssymb}

\usepackage{amssymb}

\def\'#1{{\accent19\ifx #1i \i\else #1\fi}}

\def\be{\begin{equation}}
\def\ee{\end{equation}}
\def\bea{\begin{eqnarray}}
\def\eea{\end{eqnarray}}

\newbox\Ancha
\catcode`@=11
\newdimen\ex@
\title{  \huge Cosmology from decaying  dark energy, primordial at the  Planck scale }

\author{  \Large J. Besprosvany}

\date{\Large Instituto de F\'{\i}sica, Universidad Nacional Aut\'onoma de M\'exico,
Apartado Postal 20-364, M\'exico 01000, D. F., M\'exico }

\begin{document}

\maketitle








\jot = 1.5ex
\def\baselinestretch{1.1}
\parskip 5pt plus 1pt






\Large
\begin{abstract}
\large
 The consideration of dark energy's  quanta, required also
by thermodynamics, introduces its chemical potential into the
cosmological equations. Isolating its main contribution, we obtain
solutions with  dark energy decaying to matter or radiation. When
dominant, their energy densities  tend asymptotically to a
constant ratio, explaining today's dark energy-dark matter
coincidence, and in agreement with supernova redshift data, and a
universe-age  constraint. This also connects the Planck's and
today's scales through time. This decay may be manifested in the
highest-energy  cosmic rays, recently detected.

\end{abstract}
\vskip .5cm
 \centerline{\normalsize PACS: 98.80.Bp,  98.80.Es, 04.40.-b,
98.70.-f, 14.80.-j}

 \large
 \baselineskip 22pt \vfil\eject \noindent
Dark energy  is a    component of the universe whose
 negative pressure,  characteristic of  the quantum vacuum,
 accelerates
 its
 expansion. Evidence   for its existence
 has recently accumulated from
 independent sources as  the  supernova redshift far-distance
relation$\cite{Perlmutter},\cite{Garnavich}$,  structure
 formation$\cite{structure}$,
the microwave background radiation$\cite{microwave}$, and
lensing$\cite{lensing}$.
 The coincidence of its present energy-density scale with the
 universe's, its smallness by
122 orders of magnitude  with respect to the vacuum's natural
Planck scale,
and its origin have  remained puzzling.

The cosmological constant $\Lambda$ was originally added by
Einstein  in the application of general relativity to cosmology in
1917 in order to describe a static universe$\cite{Einstein}$,
building on a 1890s proposal  by Neumann and Seeliger, who
introduced it in a Newtonian framework for the same reasons. Its
contribution in the Einstein equations
\begin{eqnarray} \label  {equationRelativity}
R_{\mu\nu}-\frac{1}{2}g_{\mu\nu}R-\Lambda g_{\mu\nu}=8 \pi
T_{\mu\nu}
\end{eqnarray}
equilibrates gravity's attraction in a matter universe; here
$R_{\mu\nu}$ is the Ricci tensor, $g_{\mu\nu}$ the metric
tensor, which describe the geometry, and $T_{\mu\nu}$
 is the energy-momentum tensor; we use units with  the Newton, Planck, Boltzmann, and   light-speed constants
$G=\hbar=k_B=c=1,$
 except when given explicitly, as needed.
 Zeld'ovich sought to connect it to the quantum vacuum$\cite{Zel'dovich}$.
 This requires  its  reinterpretation  as a component of
$T_{\mu\nu}$ in Eq. \ref{equationRelativity}. The   vacuum energy
density   of   particle fields with mass $m\ll M_P=\frac{1}{\sqrt{G}}$ is obtained
 by summing over its modes $\bf k$:
\begin{eqnarray} \label {PlanckSpec}
   \rho_{\Lambda P}  =\frac{1}{(2 \pi)^3}\int^{M_P}d^3k
\sqrt{k^2+m^2}\simeq 3 \times 10^{114} \ {\rm \frac{GeV}{cm^3}};
\end{eqnarray}
 the
natural cutoff is the Planck-mass scale $M_P$,  the only possible
mass  conformed of $G$, $\hbar$, and $c$,
while in today's universe $\rho_{\Lambda0}\simeq 4\times  10^{-6}\
{\rm GeV/cm^3}.$  Among various explanation attempts, this
striking difference  has been attributed to a  time-changing $G$
$\cite{DiracVariable} \cite{variableConst}$.

$\rho_{\Lambda0}$ represents $\Omega_{\Lambda 0}=\rho_{\Lambda
0}/\rho_{c0}\simeq .73$ of its
  critical energy density today$\cite{ParticleData}$
$\rho_{c0}$, where in a flat universe $\rho_c$ is also the total
energy density$\cite{Peebles}$. The  rest  corresponds mainly
to matter, dark and baryonic,  the latter conforming $\Omega_{b0}\simeq
 .044$
only$\cite{ParticleData}$.
  Under the
isotropic Robertson-Walker metric
 $ds^2=dt^2-R^2(t)(dx^2+dy^2+dz^2)$,
 Eq. \ref{equationRelativity} implies the  Friedmann equation
 \begin{eqnarray}\label  {Friedmann}
H^2 = \frac{8       \pi  }{3}\rho_c= \frac{8       \pi
}{3}(\rho_\Lambda+\rho_r+\rho_m),\end{eqnarray} where the other
known energy sources, radiation  $r$ and matter $m$, have been
included, $x,y,z$ are commoving Cartesian coordinates,  $R$ is the
scale factor, depending on time $t$, as do the $\rho_i$,
   and $H =   \dot R/R$   the  Hubble parameter
(a dot denotes time derivative).
  Each component $i$ has  pressure $p_i$  and is characterised by an equation of state
\begin{eqnarray} \label  {equationofstate} p_i=w_i \rho_i,
\end{eqnarray}
where  $w_r=1/3$ for radiation, and for  relativistic Fermi
or Bose gases, and $w_m=0$ for non-relativistic matter.
 The
energy-conservation equation within    an
expanding volume $V$ \begin{eqnarray}\label  {energyeqCons} \sum_i d(\rho_i V)=-\sum_i  p_i dV
\end{eqnarray}  is
implied by the contraction of Eq. \ref{equationRelativity}, and
also by thermodynamics. When decoupled,  each contribution
satisfies
\begin{eqnarray}\label  {energyeq} d(\rho_i V)=-p_i dV.
   \end{eqnarray}

  The
 form of  Eq. \ref{equationRelativity} implies  $\Lambda$ generates
 a pressure $p_\Lambda=-\rho_\Lambda$, so $w_\Lambda=-1$ for the vacuum
 energy.
The parametric extension to arbitrary negative values $w_\Lambda$, with
similar properties$\cite{Steinhardt},\cite{TurnerSmooth}$,
suits the lack of precise knowledge about it.
 Whatever is  its nature,
 and with a name not bound to its constancy,   dark energy    should contain  quanta$\cite{Zel'dovich}$, as any
other form of energy in the universe,  and so, the energy
dependence on its number $N$ should be accounted for. Eq.
\ref{energyeq} is then modified by the  chemical potential $\mu$
contribution
\begin{eqnarray}\label  {EnergyConserv}
 d (\rho V)=-p dV+\mu dN.  \end{eqnarray}
In this   letter we consider  dark-energy's
 chemical-potential modification of the cosmological equations;
its main contribution implies dark energy decays to another component.
The derived asymptotic energy-density constant  ratio of the dominant components
reproduces  the coicidence of dark energy and dark matter today, and fits
the supernova redshift data. Also, dark energy's decay
 connects Planck's scale to today's energy-density scale from Planck's time to
the universe age.

For systems satisfying Eq. \ref{equationofstate}, $\mu$
  can be obtained
 consistently with the  entropy density $s=S/V$ by extrapolating, and using the thermodynamics
relation $s=\frac{1}{T}(\rho+p-n \mu )$, where $T$ is the
temperature, and
  $n=N/V$  the particle density. When radiation-like,  $s_{r w}=c_{r
w}\rho^{\frac{1}{1+w}}$, and $\mu_{r w}=0$. Dark energy interacts
feebly, and presumably, only gravitationally; in the latter case,
the {\it single} scale in the integration constant $c_{rw}$ can
only contain the Planck scale, and an  O(1) numerical constant
(the same  for
 $c_{0w }$, $c_{w\chi}$ below.)
Such an  argument correctly gives
 $s_{r \frac{1}{3}}$, which  corresponds to the Stefan-Boltzmann law: demanding that  $c_{rw}=M_P^{3-\frac{4}{1+w}}$   not
contain $G$, as occurs for radiation, one gets $w=1/3$. If the
energy depends on $V$ through a power law, as implied by Eq.
\ref{equationofstate}, and they remain extensive, another such
quantity is required, and $N$ is the necessary choice in most
physical systems. In the zero-entropy regime $\rho_{\Lambda
w}=c_{0w }  n^{1+w}$,
and $n\mu_{\Lambda w}=(1+w)\rho .$
Non-zero temperatures or interactions may modify $n\mu_{\Lambda w}$
 to
$n\mu_{w\chi}=(1+w+\chi)\rho$, 
where $\chi$ parameterises their effect. This
 leads to
$s_{w\chi}=c_{w\chi}n(\frac{\rho}{n^{1+w}})^{-\frac{1}{\chi}},$
and
$T_{w \chi}=-\frac{\chi \rho}
{c_{w\chi} n} (\frac{\rho}{n^{
1+w}})^{{\frac{1}{\chi}}}.$
  $s_{w\chi}=s_{r w}$ for $\chi=-w-1,$  and  the zero-entropy
case is approached with $\rho\sim\rho_{w\Lambda}$, for $\chi\rightarrow 0$.
  We conclude
$\chi$ is O(1). $\chi$ need not be constant (nor $w$, for that
matter.)
 In the high-temperature regime,
a polytropic  gas has$\cite{LandauLifshitz}$ $s_{h w}=(n/w) {\rm Log
}[\rho/( c_{0w } n^{1+w})]$, for which $n\mu_{hw}=\rho\{ 1+w -{\rm Log }[
 \rho/(c_{0w }n^{1+w})]\}.$

 $\mu_\Lambda dN=
 \mu_\Lambda (n_\Lambda dV+V dn_\Lambda)$, and in the universe's evolution  in   $dt$, the partial width
 $\Gamma_1$  in
 $ N \Gamma_1 dt =n_\Lambda\mu_\Lambda   dV=(1+w_\Lambda+\chi)\rho_\Lambda dV$
 is associated with   decay due to its expansion
\begin{eqnarray}\label {Gamdos}n_\Lambda \Gamma_1=3(1+w_\Lambda+\chi) H
\rho_\Lambda \sim  \rho_\Lambda^{3/2}, \end{eqnarray} given $H\sim \rho_\Lambda^{1/2},
$ and
corresponds to changes $\partial N_\Lambda/\partial V=n_\Lambda$.
 $ N \Gamma_2 dt =\mu_\Lambda  V dn_\Lambda$ contains terms
that are not of this form. It could account for any other
conceivable decay process linked to interactions. For the
gravitational interaction, and $T=0$,  $\Gamma_2\sim \sigma
n_\Lambda v \sim (1/M_P^4) n_\Lambda\rho_\Lambda^{1/2}$, where
for the cross section $\sigma \sim(1/M_P^4) \rho^{1/2}_\Lambda$,
and the velocity $v\sim c=1$,  so
 $ n_\Lambda \Gamma_2 \sim \rho_\Lambda^{-\frac{2}{w_\Lambda+1}+1/2}$,
using $\rho_{\Lambda w}$ above.
 Therefore, for $-3<w_\Lambda<-1$,
$\Gamma_2 \ll\Gamma_1$ as $\rho_\Lambda\rightarrow 0$.
 Similarly, this will always occur for $T_{w\chi }\neq 0$,   $\sigma
\sim(1/M_P^4) T^2$, and $\rho_\Lambda\sim \rho_{w\Lambda}$.
 Another type of interaction can be dominant for some time, but it
will eventually be overridden by the $\Gamma_1$ term.
 Lower powers of $\rho_\Lambda$,  e. g.,
a  constant decay rate $n_\Lambda\Gamma_2\sim \rho_\Lambda$,
could make a
 significant cosmological
 contribution, but it  would have to be fine-tuned to give the present
 parameters.
Thus, the $\Gamma_2$ term can and will be
 neglected.

  We obtain, using Eqs. \ref{EnergyConserv},
\ref{Gamdos},
\begin{eqnarray}\label  {rholamec}
\dot \rho_\Lambda+3(w_\Lambda+1)H
\rho_\Lambda=3[(w_\Lambda+1)+\chi]H \rho_\Lambda,
\end{eqnarray}
with the latter term producing dark energy decay for $\chi< 0.$
 Energy conservation in Eq. \ref{energyeqCons}  demands that
 energy be transferred, which we assume occurs for the  $d$ component
 \begin{eqnarray}\label {RadMat}
\dot \rho_d+3(w_d+1)H \rho_d=-3[(w_\Lambda+1)+\chi ] H
\rho_\Lambda.
\end{eqnarray} The set of Eqs.  \ref{Friedmann},
 \ref{rholamec}, \ref{RadMat}
 describes $\rho_\Lambda$
as a fluid    decaying out of
equilibrium as is common in many universe
processes$\cite{Kolb}$. A  decaying cosmological constant was
first conceived by  Bronstein$\cite{Bronstein}$   to explain
the universe's time direction, and recent study starts with Ref.
\cite{Ozer}, with various phenomenological decay laws
then considered$\cite{Reuter}$.
 By substituting $H$ in   Eq. \ref{Friedmann} into   Eq. \ref{RadMat},  we  obtain
\begin{eqnarray}\label {RadMatExpli}
 \rho_d=  -{\rho}_\Lambda
 +\frac{{\dot\rho}_\Lambda^2}{ 24
     \pi  \chi^2  {\rho}_\Lambda^2}.  \end{eqnarray}
   Substituting this  into Eq. \ref{rholamec}, we get
\begin{eqnarray}\label  {MasterEq}
   6\,\chi\,\rho_\Lambda\,  \ddot{\rho}_\Lambda  + \left( d - 6\,\chi \right) \,
   {\dot\rho_\Lambda}^2
  -24\,\pi \,[ d - 3\,\left( 1 + w_\Lambda \right)
]
 \,{\chi}^2\,
   {\rho_\Lambda}^3 =0,
   \end{eqnarray}
where $d=3(w_d+1)$. $t$ as   inverse function   of
$\rho_{\Lambda}$  can be integrated, where initially $\rho_{\Lambda_i}$ at
$t_i$
\begin{eqnarray}\label  {Integrated}
t-t_i=  \int_{ \rho_ {\Lambda } }^{\rho_{\Lambda_i} }  d\rho
\left(\frac{d+3 \chi}{ 24{{\chi}}^2 \pi[d-3(w_\Lambda+1) ]
{{\rho}}^3+
 3({d+3 \chi})  {\chi \,C}{\rho}^{2 - \frac {d}{3\,\chi}}    }\right )^{\frac{1}{2}}.
   \end{eqnarray}
$C$  accounts for initial conditions for ${\rho_d}$, and we have
chosen the solution for which $R$ increases and $\rho_\Lambda$
decreases. For some $\chi,$ $w_\Lambda$, $t(\rho_\Lambda)$ can be
given explicitly in terms of hypergeometric and elliptic
functions.

Assuming $\rho_d$ is dominant, neglecting  the non-dominant term in Eq. \ref{Friedmann}, and
using Eqs. \ref{RadMatExpli}, \ref{Integrated}
  one finds
\begin{eqnarray}\label  {rhocrit}
 \rho_c\approx \frac{24{{\chi}}^2 \pi[d-3(w_\Lambda+1) ]
{{\rho_{\Lambda}}} +
 ({d+3 \chi}) 3\, {\chi \,C}{\rho_{\Lambda}}^{ - \frac {d}{3\,\chi}}}{24
     \pi  \chi^2    (d+3 \chi)}.
  \end{eqnarray}
 One derives that for $ -d/3<\chi<0
$
\begin{eqnarray}\label  {asymptotic}
{\rm lim}_{\rho_ {\Lambda} \rightarrow 0 } \frac{{\rho_
{\Lambda}}}{\rho_c} =\frac {  d+3 \chi }{d-3(  w_\Lambda+1)}
\end{eqnarray}  within the wide set of
initial conditions   $C\ll \rho_{\Lambda 0} ^{1 + \frac
{d}{3\,\chi}}$,
 so  $\Omega_d$ and $\Omega_\Lambda$ will acquire a fixed asymptotic
 value.
In this limit, the
$d$ component in  Eq. \ref{RadMat} behaves as $\rho_\Lambda$
in Eq. \ref{rholamec}\begin{eqnarray}\label {comun} \dot
\rho_d-3\chi H \rho_d=0.
\end{eqnarray}
 Such
 $\rho_d$
 and $\rho_\Lambda$
depend on the scale factor  as $R^{3\chi}$, while $R\sim t^{-2/(3
\chi)}. $ In fact, from Eqs. \ref{Integrated}, \ref{rhocrit},
\ref{asymptotic}   the dominant components produce $\rho_c\sim
1/(6 \pi \chi^2 t^2)$,
 and the representative constant $|\chi_t|=\frac{1}{t_0(6 \pi \rho_{c0})^{1/2}}\simeq .67$, using
the universe's age $t_0\simeq 13.7\pm .2$ years$\cite{ParticleData}$.
For a wide set of conditions, including matter and radiation domination, this behaviour correctly connects
today's energy densities with Planck's  scale at Planck's time
$t_P$: $\rho_0 \sim \rho_P (t_P/ t_0)^2$.

 Indeed, for $\chi\neq 0$ Eq. $\ref{rhocrit}$ implies the volume factor  $(R_f/R_i)^3$ grows
 by  $3 \frac{1}{|\chi|}{\rm Log
}(\frac{\rho_{\Lambda_i}}{\rho_{\Lambda_f}} ) $
 $e$-folds. Therefore, a small enough
  $\chi$  in the early
universe can  meet the  $10^{88}$ entropy factor that solves the
problems of {\it smoothness}, {\it flatness}, and {\it causality},
as does inflation$\cite{inflation}$.

 Eq. \ref{rholamec} with $d=4$ applies  to the  radiation-dominated
epoch. In general, $\chi_r\sim -4/3$ gives $\Omega_\Lambda\ll 1$
so as not to interfere with nucleosynthesis$\cite{nucleosyn}$,
maintaining $\eta=\rho_b/s_r$;
radiation and matter grow as in
Eq. \ref{energyeq},  $\rho_b\sim 1/R^3$, and $\rho_r\sim 1/R^4$,
as does dark matter $\rho_{dm}\sim 1/R^3$, until dark energy
becomes dominant, and  the decaying term in  Eq. \ref{rholamec}
becomes important. If dark matter interacts  weakly, dark energy's
decay's switch to dark matter occurs at $T\sim 10$ GeV, before the
onset of nucleosynthesis.
  In a second scenario,
dark energy decays simultaneously to radiation and matter, in a
proportion that is reconciled with a constant $\eta$.

 The cosmic microwave background radiation data is
consistent with a dark-energy component and
the presence of both dark energy and dark
matter provides a better fit in
structure-formation models$\cite{structure}$.
If after nucloesynthesis $\rho_\Lambda$ is small, it should  decrease slower than
radiation and matter so $\chi > -4/3$, and when matter
dominates, $\chi > -1.$ Then it   can  influence the
 structure formation, and,  eventually,  dominate together with dark matter.

Such a behavior fits
 the supernova
data$\cite{Supernova}$  interpreted under Eq. \ref{asymptotic}, with dark matter and dark
energy evolving with a constant ratio. With the simplest assumption
of  constant $\chi_0=-.48$, and as shown in Fig. 1 (which also includes the non-fitting
$\Omega_{\Lambda 0}=0$, $\Omega_{\Lambda 0}=-1$  cases),
one can reproduce the luminosity
distance  $d_L =  H^{-1}_ 0 (1 + z) \int ^ z _0 dz^\prime
[\Omega_{b0}(1 + z^\prime)^3+ (1 - \Omega_{b0})(1 +
z^\prime)^{-3\chi_0}]^{-1/2}$ up to the measured redshift $z\sim
2$, where the fit is independent of $\Omega_{\Lambda 0}$  (which  fixes $w_\Lambda$.).
This is in
accordance with a slower growth than $\rho_r$, $\rho_b$ into the
past, including the time of dark-energy-baryon equality at $z_{\Lambda b
}=(\Omega_{\Lambda 0}/\Omega_{b 0})^{\frac{1}{3(1+\chi_0)}}-1\simeq
5$, after which the asymptotic behaviour of Eq. \ref{asymptotic}
ensues. Also consistently, the time since $z_{\Lambda b }$  does not
saturate the age of the universe: $  \int^{z_{\Lambda b
}}_0 dz^\prime (1 + z^\prime)^{-1}[\Omega_{b0}(1 + z^\prime)^3+ (1
- \Omega_{b0})(1 + z^\prime)^{-3\chi_0}]^{-1/2}\simeq .9  <
 H_0 t_0\simeq 1$, and  $ \chi_r<\chi_t <\chi_0$.

Energy injection to the universe through   dark-energy decay may
have observable consequences. Most    cosmic rays can be
attributed to galactic origin, up to $ E_{cr}\simeq 10^{11}$ GeV$\cite{cosmic}$.
The  large flux and apparent
 isotropy of recently detected$\cite{beyondGZP}$
rays   with energies beyond the  GZP limit$\cite{GZP1}$  make
them difficult to associate with extra-galactic sources, and
 yet, to galactic processes$\cite{cosmic}$.
 Rays at $E_{cr}$ arrive with an
emissivity of $10^{-9} \  {\rm \frac{GeV}{cm ^2 sec} }$,  
representing $\rho_{cr}\simeq 4\times  10^{-19}{\rm \frac{GeV}{cm ^3 } }$. The  gravitational
decay of dark energy mediated by a particle of mass $m_a$ (and involving a photon or a nucleon),
can have a width $\Gamma_a\sim m_a^5/M_P^4$, and it could
 conform the energy-transfer mechanism   in the $E_{cr}$ channel with
  $(E_{cr}/M_P)^3\rho_{\Lambda 0} H_0=\Gamma_a  \rho_{cr}$;
we used  the phase-space factor in Eq. \ref{PlanckSpec}, assuming
  that  the vacuum Planck spectrum is uniformly depleted,
as suggested by relatively constant spectrum, and the prevalence
of physical constants. We find $m_a\sim 10^4$ GeV, a range for
future accelerators.

\begin{figure}[h]
 \begin{center}
 \includegraphics[scale=0.9]{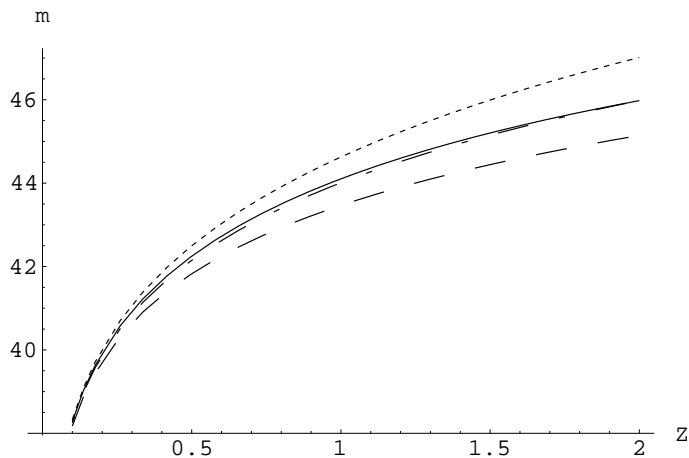}
 \end{center}
 \caption{\large Comparison of magnitude $\mu=5 {\rm Log}_{10} (d_L)+25$ of luminosity distance $d_L$,
as a function of redshift $z$, for flat models. For non-asymptotic
models with $w_\Lambda=-1$, and (a) $\Omega_{m0}=0$,
$\Omega_{\Lambda 0}=1$ (dotted), (b) $\Omega_{m 0}=.27$,
$\Omega_{\Lambda 0}=.73$
 (line),
and (c) $\Omega_{m0}=1$, $\Omega_{\Lambda 0}=0$ (dashed); and (d)
for asymptotic model with  $\Omega_{b 0}=.044,$  and $\chi=-.48$
(dot-dashed).}\label{algun--nombre}
 \end{figure}


In  summary,
account of  dark energy's quanta connects today's energy-density
scale with Planck's,  within classical general relativity and
thermodynamics. It represents a departure from the
zero-temperature cosmological constant, while it maintains
 the results of the standard cosmology.
 This supports
 a conservative approach   in   which known physical elements  can provide new
  information$\cite{JaimeLett}$.
Dark energy's coincidence  with the critical density today  is
connected to the universe evolution,
in which events occur by contingency, rather than
chance. While microphysics$\cite{Adler}$ needs
to elucidate
 the dark energy's
 equation of state,
the universe  already emerges as
 flat,
 interconnected, evolving deterministically, and  in an inexorable process of
 accelerated expansion and decay.

 \vskip 1cm

\baselineskip 22pt\vfil\eject \noindent




{\bf Acknowledgments}

 I thank  A. de la Macorra for introducing me into the subject, and acknowledge  current support
from CONACYT, project 42026-F, and from DGAPA-UNAM, project
IN120602.

\setcounter{equation}{0}


\end{document}